\documentclass{nature}
\usepackage{color}
\usepackage{arydshln}
\usepackage{graphicx}  
\usepackage{lineno}
\usepackage{booktabs}  
\usepackage{threeparttable}
\usepackage[final]{pdfpages}
\usepackage{amsmath, amsthm, amssymb}
\usepackage{bm}
\usepackage{multicol}  
\usepackage{multirow}

\makeatletter
\let\saved@includegraphics\includegraphics
\AtBeginDocument{\let\includegraphics\saved@includegraphics}
\renewenvironment*{figure}{\@float{figure}}{\end@float}
\makeatother

\title{Formation of intermediate-mass planets via magnetically-controlled disk fragmentation}
\author{Hongping Deng$^{1,  2, *}$, Lucio Mayer$^2$, Ravit Helled$^2$}

\begin{document}

\maketitle

\begin{affiliations}
\item Department of Applied Mathematics and Theoretical Physics, Centre for Mathematical Sciences, University of Cambridge, Wilberforce Rd, Cambridge CB3 0WA
 \item Center for theoretical Astrophysics and Cosmology, Institute for Computational Science, University of Zurich, Winterthurerstrasse 190, 8057 Zurich, Switzerland
\end{affiliations}

\begin{abstract}
Intermediate mass planets, from Super-Earth to Neptune-sized bodies, are
the most
common type of planets in the galaxy \cite{Schneider2011}.
The prevailing theory of planet formation, core-accretion
\cite{Pollack1996}, predicts significantly fewer intermediate-mass giant planets than observed
\cite{Mordasini2018,Suzuki2018}.
The competing mechanism for planet formation, disk instability, can
produce massive gas giant planets on wide-orbits, such as HR
8799\cite{Marois2008}, by direct fragmentation of the protoplanetary disk\cite{Boley2010}. Previously, fragmentation in  magnetized protoplanetary disks has only been
considered when the magneto-rotational
instability is the driving mechanism for magnetic field growth\cite{Fromang2005}. Yet,
this instability is naturally superseded by the spiral-driven dynamo
when more realistic, non-ideal MHD conditions are considered \cite{Riols2019, Deng2020}.
Here we report on MHD
simulations of
disk fragmentation in the presence of a  spiral-driven dynamo. 
Fragmentation leads to the formation of
long-lived
bound protoplanets with masses that are at least one order of magnitude
smaller than in conventional disk instability models\cite{Boss2003a,Nayakshin2017}. These light clumps survive shear and do
not grow further due to the shielding effect of the magnetic field,
whereby magnetic pressure stifles local inflow of matter. The outcome is a
population of gaseous-rich planets with intermediate masses, while gas giants are
found to be rarer, in qualitative agreement with the observed mass
distribution of exoplanets.
\end{abstract}

We perform 3D MHD simulations with
self-gravity, as well as, for comparison,  equivalent self-gravitating
disk simulations without a magnetic field.
We model a 0.07$M_\odot$ protoplanetary disk spanning 5-25 AU around a solar mass ($M_\odot$) star starting from a quasi-equilibrium state of \cite{Deng2020}. 
Such a disk mass, close to $10\%$ of the mass of the central star, is required for self-gravity to have
an important dynamical impact, and is expected to be common in the early stages of disk evolution, being
a product of the initial cloud collapse phase \cite{Li2014}.
We use a novel meshless finite mass scheme\cite{Hopkins2015,Hopkins2016} which is capable of capturing
the complex regime of magnetized self-gravitating flows \cite{Deng2019,Deng2020} (see Methods).  
Although fragmentation
conditions would be more easily met in a more extended disk, a compact
disk model allows to maximize the number of resolution elements per
vertical scale height, and to capture correctly MHD effects
\cite{Deng2019}.
In order to probe the low-mass-end of fragments we use a mass resolution of $~2\times 10^{-6}M_J$ in the MHD simulation. 
The computational cost of the extremely high resolution prevents the use of radiative transfer. Hence we adopt a standard parameterized cooling prescription\cite{Gammie2001}. The cooling  time scale $\tau_c$ is governed by the dimensionless parameter $\beta_c=\tau_{c}\Omega$ where $\Omega$ is the radial dependent Keplerian rotational frequency. Our two stage simulations apply $\beta_c=3$ to trigger fragmentation\cite{Gammie2001,Deng2017} and $\beta_c=6.28$ to trace the subsequent evolution of clumps in a gravito-turbulent disk (see Methods).   
Fast cooling as with $\beta_c=3$ should occur only in the outer regions of protoplanetary disks, but also mass loading due to infall can promote
fragmentation if it is fast enough \cite{Boley2009}. 
We start the simulation with $\beta_c=3$ because we focus on the effect of the 
magnetic field on fragmentation provided that favourable conditions hold
for fragmentation to take place.

\begin{figure}
\centering
  \includegraphics[width=0.95\textwidth]{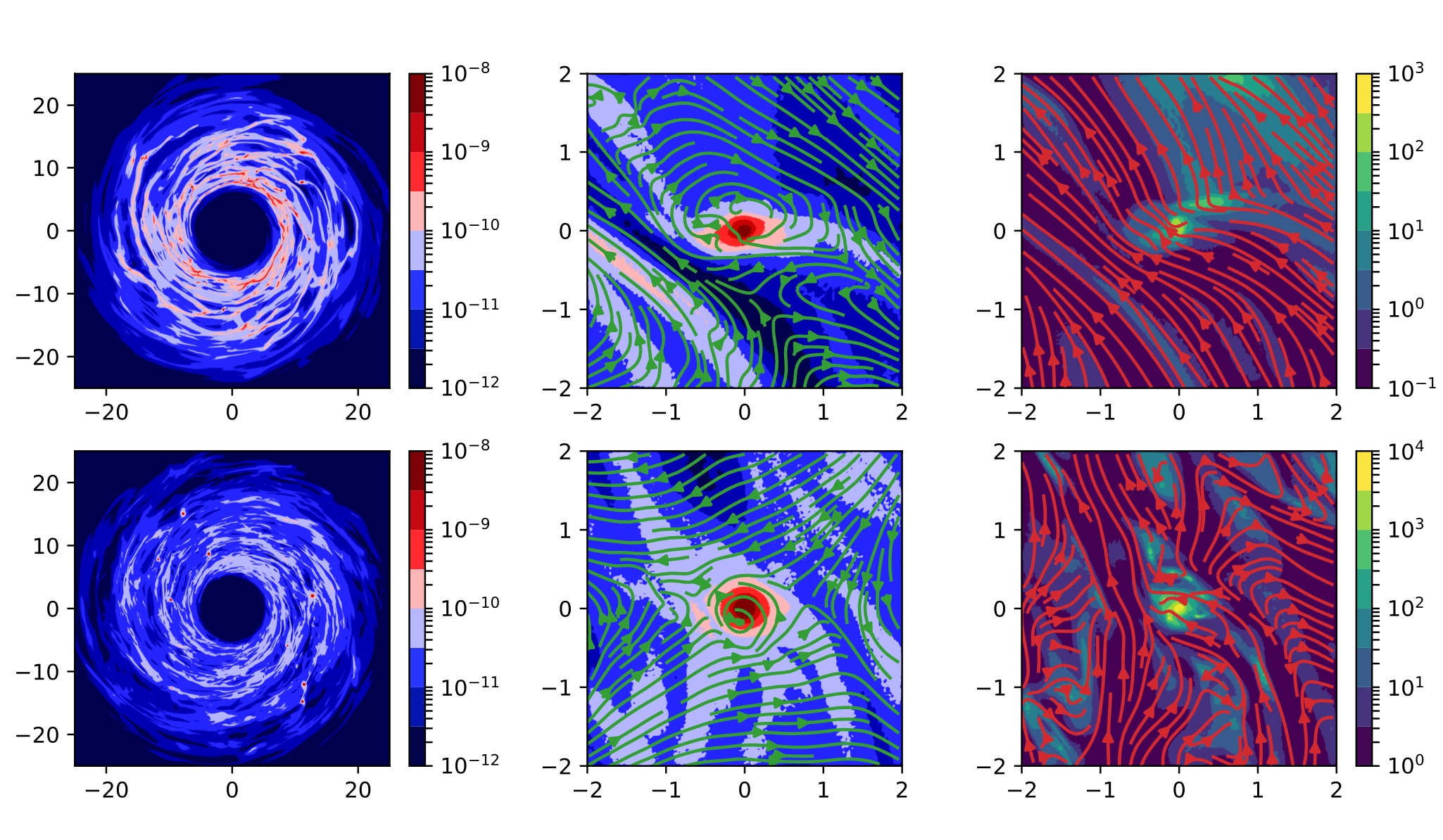}
  \caption{Global and local flow and magnetic field structures in the disk midplane. {\bf Upper panels:} The MHD turbulent disk just before we switch the cooling parameter $\beta_c$ from 3 to 6.28 at 100 yrs. Several gravitationally bound clumps are formed as indicated by the midplane density (in g/cm$^{-3}$) in the upper left panel. The upper middle panel zoom in on the clump at 2 o'clock showing the density field with over-plotted velocity streamlines. The upper right panel shows the ratio between gas pressure and magnetic energy, i.e., $\beta_p$ with magnetic fields over-plotted. {\bf Lower panels:} Same as the upper panels but for the disk at t=290 yrs. The zoomed-in lower panels focus on the clump at 3 o'clock of the lower left panel. we note here the length unit is 1 au.}
\end{figure}

First of all, we find that $\beta_c=3$  is sufficient to trigger fragmentation
in the MHD simulations. The fragmentation simulations were run for 100 yrs (about one outer rotation period). 
In the MHD simulation, four protoplanets (gravitationally bound regions) of 0.01-0.02$M_J$ and five even smaller protoplanets form from condensation of 
material  in the spiral arms (Figure 1, upper left). Two clumps in the equivalent HD simulation, HD-cl1 and HD-cl2 (Extended data figure 1) 
%form hosting 
are formed, with protoplanet masses of $0.128M_J$ and $0.042M_J$. 
The MHD simulation leads to more clumps due to the different power spectrum of the density fluctuations relative to the HD simulations, which
is skewed to higher order modes with short wavelength \cite{Deng2020}. 
The clumps, despite their small masses, begin to stir rotational flow around them locally, and the magnetic fields aligned with the spiral arms \cite{Riols2019} are curved and amplified by the flow (Figure 1). In this initial phase
the magnetic energy is less than 1\% of the gas pressure (plasma beta $\beta_p>100$) in the inner region of the clumps (Figure 1, upper right). As a result, the collapse of spiral arm and the early growth of protoplanets is hardly affected by the magnetic fields (Figure 2, Time $<$100 yrs).

The disk keeps fragmenting in over-dense regions if we keep applying such a fast cooling. However, this is unrealistic as the dense gas in the
dense spiral arms would become rapidly optically thick, which in turn reduces fragmentation and mass accumulation by already existing clumps\cite{Mayer2007,Szulagyi2017}. 
%A conventional solution is to introduce sink particles \cite{Bate1995}, (\textbf{The logic doesn't, sink particle is used to enable long term simulation} this is known to induce numerical
%artifacts and severely overestimate the number of fragments and their masses compared to simulations with radiative transfer (Durisen et al. 2007).
After 100 years, namely after the first orbit on which fragmentation happens, we thus both reduce the
cooling rate by a factor of two and shut-off cooling above a critical density of $10^{-9}$g cm$^{-3}$ to account for the diminishing 
radiative losses at higher optical depths, as suggested by simulations with radiative transfer  (see Methods).
%The density threshold is chosen such that above it the optical
%depth is large enough for the gas to suffer negligible radiative losses (see Methods), hence behaving nearly adiabatically\cite{Mayer2004}. 
We focus on the evolution of low-mass clumps, especially on how the strong magnetic fields ($\beta_p\sim 1$) around it affect its evolution (Figure 1). When we switch $\beta_c$ to 6.28 at t=100 yrs, low-mass fragments are formed. We can
follow their long-term evolution in a gravito-turbulent disk, and compare between the HD and MHD simulations.

Without continued efficient cooling in the inner region (Extended data figure 2), the light protoplanets within the HD simulations  are quickly disrupted by
shear\cite{Forgan2017} (their Figure 5). We verified that this still occurs even when we increase the resolution by
a factor of 4 (see Extended Data Figure 3).
On the contrary, 
in the MHD simulations
small  protoplanets  remain gravitationally bound and grow to roughly Neptune mass, 
{\bf }The fragments, by perturbing the surrounding flow, trigger the formation of  additional even lower
mass objects \cite{Meru2015} (Figure 2, MHD-cl3).
The remarkable difference between the HD and MHD runs is caused by the ability of the magnetic field to control the local 
flow around the clumps. The lower panels of figure 1 show a snapshot of the MHD evolution simulation at 290 yrs.

Zoom-in around MHD-cl1 reveals turbulent magnetic fields (Figure 1). 
In regions intermediately outside of the protoplanet's Hill sphere, magnetic pressure does not only dominate over the gas pressure but it also dominates 
over the kinetic energy (Extended data figure 4). When clumps form the fluid settles in rapid rotation around their center. 
The magnetic
field in the clump's region is then amplified efficiently as it twists and folds due to fluid rotation, 
despite the inclusion of Ohmic dissipation (see Methods).
Such strong fields near the Hill sphere act to isolate the clump from the disk environment. 
As a result, the material exchange rate across the Hill sphere of a Neptune mass protoplanet is significantly lower relative to the non-magnetized case
(Extended data figure 5).  In order to further investigate the role of MHD turbulence on low-mass clump accretion, we turned off the MHD module after 400 years (Figure 2), and restarted the simulation. 
We find that, after only 50 years, all clumps are already $\sim$10\% lighter than their counterparts in the standard MHD simulation, reflecting the missing shielding action of the magnetic field. 

\begin{figure}
  \centering
  \includegraphics[width=0.8\textwidth]{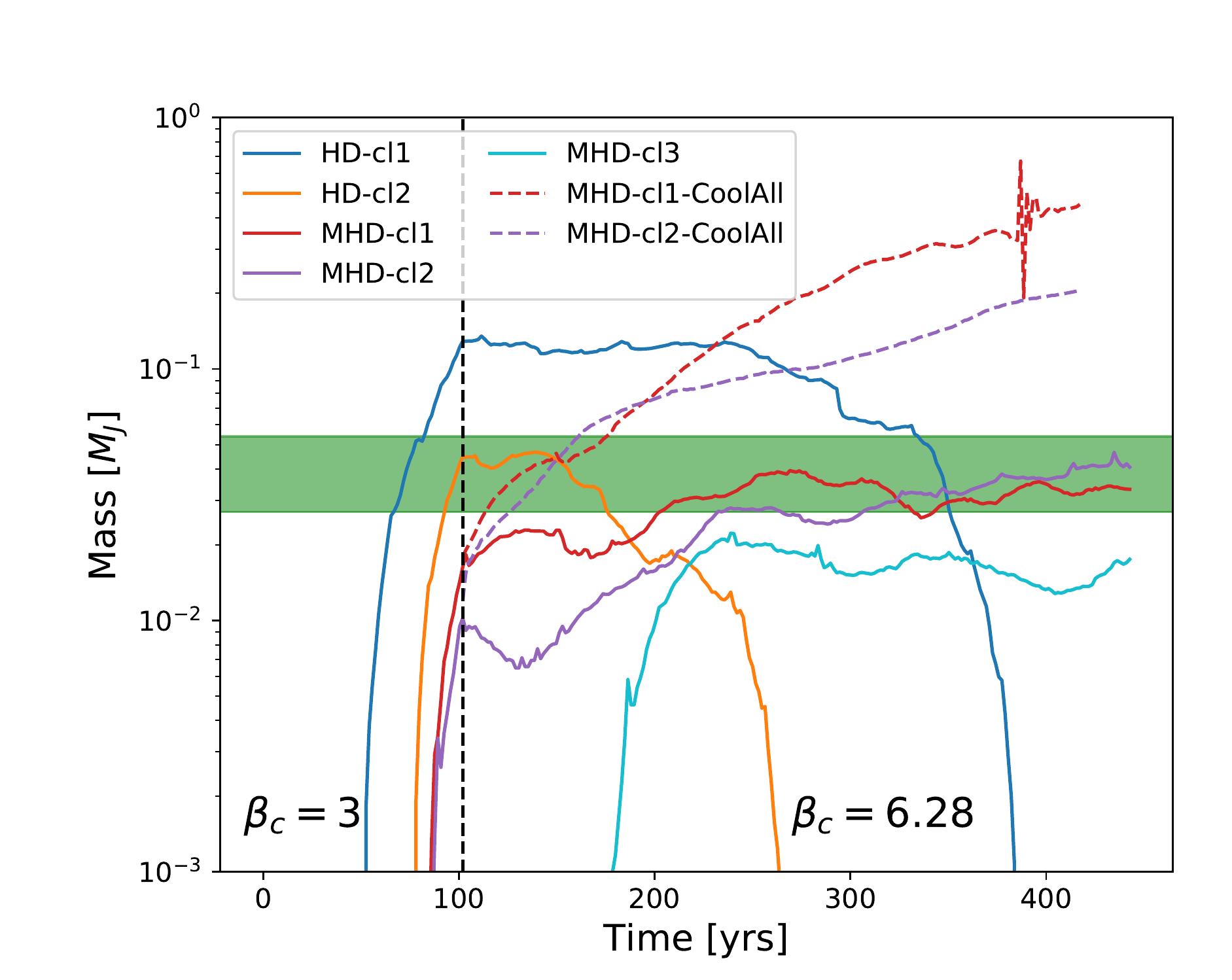}
  \caption{The mass evolution of protoplanets identified via their gravitationally bound components. We show two representative clumps in the HD simulation and two representative clumps in the MHD simulation (see Methods, subsection 2). The vertical dashed line separates the fragmentation simulation with $\beta=3$ and the later evolution with a slower cooling $\beta_c=6.28$.  The two protoplanets formed in the HD fragmentation simulation are destroyed quickly after we switch to $\beta_c=6.28$ and turning off cooling in clump core regions. On the contrary, two light protoplanets, $0.018M_J$ in MHD-cl1 and 0.010$M_J$ in MHD-cl2, in the MHD fragmentation simulation survive and grow to 0.035$M_J$ and 0.042$M_J$ at 450 yrs lying in the mass range of 0.5-1 Neptune mass (shaded green region). One protoplanet (cyan curve) is formed in the dense  region in the MHD fragmentation simulation. When we apply $\beta_c=6.28$ to all regions in the MHD evolution simulation, the protoplanet in MHD-cl1-CoolAll and MHD-cl2-CoolAll grow to 0.42$M_J$ and 0.22$M_J$ after 420 yrs. The wiggles of the red-dashed curve are caused by clump-clump mergers. }
\end{figure}

The MHD turbulence still allows growth of the very light fragments in the long-term. 
The 0.5-1 Neptune mass range appears to be a sweet-spot as three out of six protoplanets are within this mass range, and the other three clumps have masses of $0.009M_J$, $0.018M_J$ and $0.015M_J$ at the end of the main MHD simulation. This is shown by the mass function in Figure 3. The Neptune-mass protoplanets arising at distances of 
10-20 AU experience chaotic radial migration (Extended data figure 6), suggesting that
the emerging population of exoplanets is expected to have a wide range of distances from the
central star.

%to for can 
%Their composition can 
%Dust are expected to settle in these clumps forming planetary cores later on.  
The exact number of clumps and their final mass could still vary depending  the nature of the triggering perturbation, such as infall or cooling rate variations
(see Methods), and on the details of thermodynamics of the local flow near spiral arms.  
If we allow cooling at all densities (unphysical), thus maintaining $\beta_c=6.28$ at the clump's center (all particles) in the main MHD simulation, the protoplanets grow much faster and reach a few tenth of a Jupiter mass in a comparable simulation time (Figure 2). Yet, the combination of the small initial fragmentation
scale and the shielding of low-mass clumps by the magnetic field are unique features of magnetically-controlled fragmentation that naturally 
favour the formation of a population of sub-Jovian mass planets. 

Since Neptune-mass planets are typically heavy-element dominated, in order to explain their origin with this new formation path should be accompanied by heavy-element accretion in the form of  pebbles or planetesimals, and core formation via the settling of the heavy-element material 
towards the center. Core formation might render these
relatively light planets further resilient to disruption by shear and tidal effects, which, if anything, would act to reduce their mass further\cite{Nayakshin2017}.
Depending on stochastic local variations
in the  dynamics of the
turbulent flow, a minority of the clumps is observed to grow faster than
average, and could become a gas giant (e.g., Figure 2 MHD-cl2).

\begin{figure}
  \centering
  \includegraphics[width=0.8\textwidth]{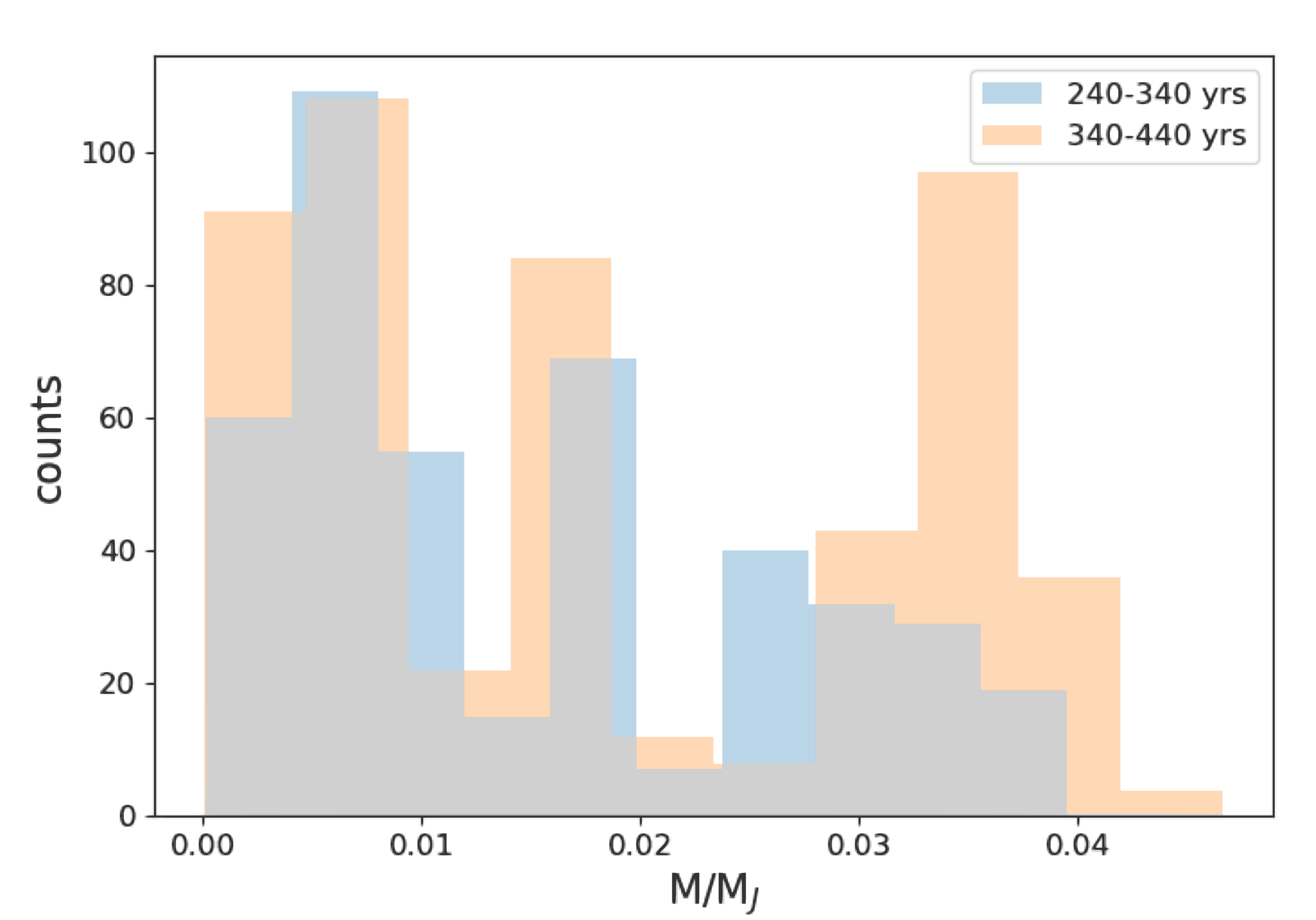}
  \caption{The frequency of planets of different masses. The planets are identified in 60 consecutive snapshots taken between 240-340 yrs and 340-440 yrs of the main MHD simulation. The mass within one sampling kernel is 0.004M$_J$. It is clear that the masses of the protoplanets do not evolve monotonically (Figure 2).}
\end{figure}

The heavy-element mass that can be accreted by the clump at a given radial distance $a$ is given by\cite{2006Icar..185...64H}  
$dm(a)/dt = \pi R_{cap}^2 \sigma (a,t) \Omega (a)$,
where $R_{cap}$ is the capture radius,  $\sigma$  is the surface density of solid material and $\Omega$ is the clump's orbital frequency. In this formula the gravitational enhancement factor is assumed to be unity. 
The captured mass would depend on the pre-collapse timescale of the clump and whether it migrates. Accurate simulations of the long-term orbital and thermal evolution 
are clearly required. 
Nevertheless, if our Neptune-mass clump located at 10.5 AU has a pre-collapse timescale of $\sim$1 Myr, as expected for small objects\cite{2010Icar..207..503H,2008Icar..195..863H}, it could accrete at least a few $M_{\oplus}$ of heavy elements in the form of planetesimals. 
%but given the uncertainty in the parameters the mass can vary from less than 1 $M_{\oplus}$ to nearly ten. 
%Since the clump's internal temperatures are very low ($<$ 100 K), the heavy elements remain solid and are expected to settle to the center to form a core\cite{2008Icar..195..863H}. 
If the heavy elements are in the form of small dust grains that are coupled to the gas when the clumps are formed, the situation is somewhat different, although planetesimal accretion could take place at a later time and enrich the planets with heavy elements. In addition, dust grains (sizes of 10cm -- 100m) are expected to accumulate around spiral arms via dust trapping and lead to an enrichment of the clumps ``from birth”, where the enrichment was estimated to be about two times stellar\cite{2010ApJ...724..618B}.  
If most of the heavy elements are in the form of pebbles, they could also be accreted after the clump formation. Pebble accretion by small clumps was found to be very efficient and can also be of the order of tens of $M_{\oplus}$\cite{2018MNRAS.477..593H}. 
The exact accreted mass of the heavy elements does not only depend on the sizes of the accreted solids and when they are accreted, but also on the disk's metallicity and surface density.  
Overall, we suggest that small clumps formed by GI could have a large range of compositions and final masses, which is consistent with the diversity of planets in this mass-range. %\textbf{Our suggested model does not  require post-formation  mechanisms to decrease  
%the mass of fragments such as radiative feedback during collapse\cite{Stamatellos2015}, photoevaporation\cite{Boss2003a} or
%tidal mass loss\cite{Boley2010, Nayakshin2017}, which  depend on quite specific environmental
%conditions\cite{Helled2014}.} 

%The fate of such clumps is still unknown, these clumps could migrate inward and accerte much mass to become gaseous planets, or dissolve back into the disk.  

We find that in magnetized gravitationally unstable disks the formation of massive gaseous planets  is less frequent in comparison to intermediate-mass planets, in agreement with the observed exoplanet population. Clearly, the diversity in disks' masses, compositions, and lifetimes are expected to promote a large spread in the inferred 
planetary properties.  
Our suggested model does not  require post-formation  mechanisms to decrease  
the mass of fragments such as radiative feedback during collapse\cite{Stamatellos2015}, photoevaporation\cite{Boss2003a} or
tidal mass loss\cite{Boley2010, Nayakshin2017}, which  depend on quite specific environmental
conditions\cite{Helled2014}.
In addition, the formation of clumps at early times, along with their dynamical
feedback on dust and pebbles, could change the properties of protoplanetary disks and influence the conditions for subsequent planet formation episodes by any mechanism, including core-accretion.

The early formation of sub-Jovian mass planets
could explain the origin of the dust rings
in the young protoplanetary  disks. 
%According to models
For example, the observed rings in IR63 in the Ophiucus molecular cloud could be triggered by embedded planets with masses less than half of that of Jupiter\cite{Segura2020}.
The formation of such  planets in a disk of age $<$ 5$\times 10^5$ years is unexpected in the standard core-accretion scenario which typically requires longer formation timescales. In the outer regions of the disk, this is also unlikely in 
the standard disk instability model, as it requires post-formation processes,
such as tidal downsizing, that are effective only  when the
clumps migrate to the inner few AU.
Therefore, the direct rapid formation of gaseous sub-Jovian mass planets can change our entire view of the process of planet formation.

%of gas giants, 
%been invoked to reduce their final masses even further.
%The impact of these effects is still uncertain
%because they depend on a large number of environmental
%conditions\cite{Helled2014}.
%as well as the early stages of disk evolution. 

%Our study demonstrates the inaccurate mass estimate for protoplanets formed by GI. 
%based on the assumption that these fragments engulf all materials within half Toomre wavelength immediately upon fragmentation. 
%We show that clump accretion and their further growth 
%of protoplanets 
%can be very inefficient for low-mass clumps. We find that MHD turbulence has only a small effect during the  initial clump collapsing stage but tends to control the subsequent clump accretion and formation of protoplanets. 
%We show that direct formation of Neptune-mass planets at 10--20 AU  is possible through fragmentation of a magnetised gravito-turbulent disk. 

%\bibliographystylelatex{naturemag}
%\bibliographylatex{reference}

\begin{methods}

  \subsection{Numerical method and initial conditions}
We simulated gravitationally  unstable protoplanetary disks using the meshless finite mass (MFM) scheme in the GIZMO code\cite{Hopkins2015,Hopkins2016}. We used the HLLD Riemann solver in GIZMO, equivalent to a second order Godunov method, and the Wendland C4 kernel with 200 neighbouring particles. MFM, as a Lagrangian method, demonstrated excellent conservation property in GI disk simulations \cite{Deng2017} and can capture subsonic MHD turbulence in protoplanetary disks reasonably well \cite{Deng2019}. The equations we solve are those of compressible self-gravitating MHD:
\begin{align}
  \frac{\partial \rho}{\partial t}+&\bm{\nabla}\cdot(\rho \bm{v})=0,\\
  \frac{\partial \bm{v}}{\partial t}+\bm{v}\cdot\bm{\nabla}\bm{v}&=-\frac{1}{\rho}\bm{\nabla}(P+\frac{B^{2}}{8\pi})+\frac{(\bm{B}\cdot\bm{\nabla})\bm{B}}{4\pi\rho} -\bm{\nabla}\Phi, \label{eq:eom} \\
  \frac{\partial \bm{B}}{\partial t}&=\bm{\nabla}\times(\bm{v}\times\bm{B})+\eta \bm{\nabla}^{2}\bm{B},  \\
  \frac{\partial U}{\partial t}& +\bm{\nabla}\cdot (U \bm{v})=-P\bm{\nabla}\cdot\bm{v}-\frac{U}{\tau_c},
\end{align}
where  $\rho$, $U$, $P$, and $\bm{v}$ are the density, internal energy, gas pressure, and velocity respectively. Here $\Phi$ is the sum of the gravitational potential of the central object and the gravitational potential induced by the disk itself, $\Phi_s$, which satisfies the Poisson equation
\begin{equation}
\nabla^{2}\Phi_{s}=4\pi G \rho.
\end{equation}
 $\bm{B}$ is the magnetic field and $\eta$ is the magnetic resistivity which is described by the dimensionless parameter $R_m=c_s H/\eta$ where $c_s, H$ are the sound speed and disk scale height.

We assume an ideal gas equation of state (EOS),
\begin{equation}
P=(\gamma -1)U,
\end{equation}
with $\gamma=5/3$. The \emph{ad hoc} cooling time scale can be parametrized by the local dynamical time scale, $\beta_c =\tau_c\Omega(r)$ \cite{Gammie2001}. We performed a two stage simulation using $\beta_c=3$ to trigger fragmentation\cite{Deng2017} and then switch to a slower cooling with $\beta_c=6.28$ to follow fragmentation evolution (Table 1).

In order to limit the computational burden in preparing the gravito-turbulenct disk model \cite{Deng2017}, we build initial conditions from the saturated gravito-turbulent disks at a cooling rate $\beta_c=6.28$  from \cite{Deng2020}. Both the HD and MHD disk model simulate a solar mass ($M_\odot$) star with a protoplanetary disk of ~0.07$M_\odot$ spanning from 5-25 au (the surface density is inversely proportional to the heliocentric distance). We note that this choice of the disk mass is conservative, as most published disk instability
simulations employ disks with masses $0.1-0.15 M_{\odot}$ around solar mass stars, which are really at the very high mass end of the observed
distribution of disk masses \cite{Helled2014}.  We split the particles once in the HD gravito-turbulent disk resulting in a mass resolution of  $\sim 2\times 10^{-5}M_J$ (3.5M particles in total) so that a Neptune mass protoplanet is resolved by $ \sim2500$ particles.  The MHD simulation employs ten times more particles to resolve subsonic MHD turbulence\cite{Deng2020}. We used much more particles than previous MHD simulations of self-gravitating protoplanetary disk of \cite{Forgan2017}. The low resolution simulation of \cite{Forgan2017} with simple initial field geometry is hard to interpret due to potential numerical field growth \cite{Deng2019}. In order to verify that the markedly different clump evolution between the MHD and HD runs is not due to the different resolution, we also performed a test HD simulation with four times more particles and found again clump disruption after a  similar time as in the default run  (Extended data figure 3). This confirms that is the action of the magnetic field to induce a diverging evolution between HD and MHD runs. 
As a final remark, we note that for disks with higher  masses as often used in disk instability simulations one expects some clumps to survive even without magnetic field. However,  these
clumps are very massive, ranging from a few times  Jupiter's mass to masses of  brown dwarfs \cite{Helled2014,Stamatellos2015, Szulagyi2017}, unless additional
mass-loss mechanisms are invoked\cite{Boley2010, Nayakshin2017}. Future MHD simulations should explore 
the mass function of clumps in disks of various masses, including disks more massive than those considered here.

%f resulting from the combined action of the magnetic field.}

The ionization rate and thus the strength of non-ideal MHD effects are highly uncertain \cite{Flock2015}. As the first attempt, we apply $R_m=20$ throughout the disk because full non-ideal MHD with coupled ionization chemistry is still challenging even for non-self-gravitating disk \cite{Wang2019}. Nevertheless, we note that the large scale dynamo in gravito-turbulent disk thrives at low $R_m$ in contrast to MRI dynamo \cite{Riols2019}. In the meantime, the fields are amplified by the local circulation near the Hill sphere (Fig. 1) on a time scale $t_{amp}=1/\Omega_H$, where $\Omega_H$ is the rotational angular speed around the clump center.  The fields near the Hill sphere can diffuse across one disk scale height (comparable to the Hill radius) in $t_{diff}=H^2/\eta=R_m/\Omega$, where  $\Omega$ is the orbital angular speed around the star. In our models, $t_{amp}=t_{diff}/(\sqrt{3}R_m)\sim 0.03t_{diff}$ so the fields are  amplified faster than they can be dissipated near the Hill sphere, hence their action remains effective in isolating the clump from the surrounding disk flow. The initial gravito-turbulent ideal-MHD disk model drawn from \cite{Deng2020} is relaxed with Ohmic dissipation switched on ($R_m=20,\beta_c=6.28$) for 100 years reaching a quasi-equilibrium state which serves as the initial condition for run MHD in Table 1. We note that the MHD simulation has zero net flux.
  
\subsection{Thermodynamics of simulations}
To trigger fragmentation of the gravito-turbulent disks, we reduce $\beta_c$ to 3 \cite{Gammie2001,Deng2017}. We note that this is just a convenient way
  to achieve the desired fragmentation conditions, but in the real astrophysical context there are additional phenomena that could bring to the
  same outcome.   Indeed, infall of gas from the molecular
  cloud envelope  can also decrease the Toomre Q parameter rapidly and lead to fragmentation, as shown in previous simulations \cite{Mayer2004, Boley2009}. In our case an infall rate of $1\times 10^{-4}M_\odot$/yr,
  expected in the early stages of disk evolution\cite{Li2014}, would induce a sudden drop of the Toomre parameter on a timescale as small as the cooling timescale
  obtained by decreasing   $\beta_c$ from 6.28 (initial equilibrium) to 3. The fast cooling leads to collapse of spiral arms (Fig. 1). We refer to over dense regions within which gravitationally bound region forms as clumps. We identify gravitationally bound regions (protoplanets) using the friend-of-friend group finder SKID \cite{Stadel2001, Backus2016}. In general, the HD simulation gives rise to fewer clumps than the MHD simulation because of more prominent spiral arms with smaller azimuthal wavenumber\cite{Deng2020}.

\begin{table} 
\caption{Simulations performed}
\label{tab:1}       % Give a unique label
\small
\begin{tabular}{ccccccc}  
    \toprule
    \multirow{2}{*}{Run}&  
    \multicolumn{3}{c}{Stage 1 (0 to 100 years)} &\multicolumn{3}{c}{Stage 2 (100 to $>$420 years)}\cr  
    \cmidrule(lr){2-4} \cmidrule(lr){5-7}  
    &Resolution&Cooling $\beta_c$&Protoplanets    &Resolution&Cooling $\beta_c$&Protoplanets\cr
    \midrule  
    HD&2E$^{-5}M_J$&3 (all $\rho$)&2&2E$^{-5}M_J$&6.28 (only $\rho<\rho_0$)&0\cr  
    MHD&2E$^{-6}M_J$&3 (all $\rho$)&9&2E$^{-6}M_J$&6.28 (only $\rho<\rho_0$)&6\cr  
    MHD-CoolAll&NA &NA &NA&2E$^{-6}M_J$&6.28 (all $\rho$)&6\cr
    \bottomrule
    \end{tabular}  
\end{table}

\clearpage
  %A more massive MHD disk (0.13$M_\odot$) with stronger spiral arms\cite{Deng2020} also forms few clumps than our fiducial model model.

 By switching $\beta_c$ back to 6.28 at t=100 yrs, we account for the suppression of radiative losses owing to the increasing optical depth
  in the dense regions of the spiral arms where clumps form. While the optical depth will vary across the disk, and especially between the
  spirals and the inter-arm regions, our simple choice produces a temperature structure at fragmentation sites that is consistent
  with the results of simulations including radiative transfer with the flux-limited diffusion approximation, the current state-of-the-art
  in the field. We refer, in particular, to the high-resolution HD disk instability simulations in \cite{Szulagyi2017}, which used a disk with a similar mass and density profile. From a clump's center outward the temperature varies typically from 200 K  to 20 K in our runs, which is very similar to the temperature profile shown in Fig.6 of the latter paper in the case of the flux-limited diffusion
  run. On the numerical side, we also note  that, if we keep $\beta_c=3$, fragmentation continues throughout the disk at a fast pace, yet forming
  clumps with similar initial masses. However, in the meantime
  the earliest forming clumps contract significantly, thus reducing the minimum time-step size and bringing the numerical integration to a bottleneck, a problem
  that is avoided by raising $\beta_c$ to $6.28$. 
   Further clump formation is suppressed after we increase $\beta_c$, yet 
  at this point we already have enough low mass protoplanet seeds for which to track the subsequent evolution.
  We tested that if increase $\beta_c$ to $6.28$ after $200$ years instead of after $100$ years the main result does not change, namely the MHD runs still have a dominant 
  population of long-lasting low mass clumps by  about 400 years, albeit in this case there are also some clumps approaching the mass of Jupiter due to higher accretion rates during the initial phase with rapid cooling.
  
  In the default HD and MHD simulations we also deliberately turn off  cooling in the interior of clumps (Extended data figure 2) where the density is above $10^{-9}$g/cm$^{3}$ (such high density is never reached in spiral density waves) to account for high opacity and slow cooling in these regions \cite{Boss2002}. We checked indeed that
  the photon diffusion timescale from the clump
  interior to the Hill radius of the clump would be
  longer than the local dynamical time.
  Assuming the chosen density threshold $\rho_0 =
  10^{-9}$g/cm$^{3}$, a typical dust opacity value
  $\kappa=1$ cm$^2$/g, and $l = 0.1$ AU for the
  path length from the clump center to its Hill
  radius, the photon diffusion timescale is
  $t_{diff} = \rho_0 \kappa l / c$, where $c$ is the
  speed of light, and must be further augmented
  by the ratio between thermal gas pressure and
  radiation pressure ($\sim 10^6$, see \cite{Boss2002,Mayer2007}).
 This is much longer than the free-fall time of the clump, also estimated at the same  threshold density, $t_{ff} \sim 1/\sqrt{(G\rho)}$,
  which is the timescale over which a clump will contract and heat-up by PdV work.
  While the latter is a simple estimate, essentially treating clump collapse as an adiabatic process, we stress that
  incorporating radiative transfer \cite{Mayer2007} into our MHD simulation would be computationally daunting as the current calculations already required 2 million CPU hours each on nearly a thousand cores on the Cray XC40/XC50 ``PizDaint" of the Swiss National Supercomputing Center (CSCS), the
  fastest European supercomputer. Furthermore, purely hydrodynamical fragmentation simulations adopting a similar cooling   shut-off threshold have obtained results qualitatively similar to radiative transfer
  simulations once clumps have formed \cite{Helled2014}.
  
 Since the photon diffusion timescale argument reported above is still simplified, neglecting for example the possibility that radiation might propagate
 anisotropically and still manage to leak out rapidly of a medium with average high optical depth, 
 we also carried out MHD simulation without turning off cooling in the clump core region, i.e., keeping $\beta_c=6.28$ for all particles (Table 1, MHD-CoolAll). The comparison MHD simulation inevitably leads to higher peak central density in the clumps than that in our main simulation. To avoid the code to crash due to the extreme dynamic range arising in this case we merge equal mass particles whose smoothing length becomes smaller than a minimum gravitational softening of 0.01 au (correspond to $>5\times 10^{-8}$ g/cm$^{3}$ regions in the very center of massive clumps. This particle merging approach results in a particles mass ceiling with the maximum particle mass forced to be 64 times the minimum particle mass. We observe no artificial effects at the boundary of different mass particles\cite{Hopkins2015}. It helps to avoid introducing a sink particle \cite{Bate1995} which artificially remove the pressure support from the clump center, potentially leading to unphysical accretion \cite{Fletcher2019}.

  \subsection{Code availability}
  The latest version of the GIZMO code is made available by its author, Philip Hopkins at http://www.tapir.caltech.edu/~phopkins/Site/GIZMO.html
  \subsection{Data availability}
  The data files that support our analysis will be made available upon reasonable request.
  
\end{methods}

%% Put the bibliography here, most people will use BiBTeX in
%% which case the environment below should be replaced with
%% the \bibliography{} command.

% \bibitem{dummy} Articles are restricted to 50 references, Letters
% to 30.
% \bibitem{dummyb} No compound references -- only one source per
% reference.
% \end{thebibliography}

%% Here is the endmatter stuff: Supplementary Info, etc.
%% Use \item's to separate, default label is "Acknowledgements"

\begin{addendum}
 \item H.D. acknowledge support from the Swiss National Science Foundation via an Early Postdoctoral Mobility Fellowship. We are grateful to Doug Lin and Gordon Ogilvie for stimulating discussions, and to the anonymous referees for constructive comments that helped to improve the manuscript significantly.
We also thank the Swiss National Supercomputing Center (CSCS) for their continued support as users of the PizDaint supercomputer, on which all the
simulations were performed.
 
 \item[Contributions] H.D. planned the study and carried out the simulations. H.D. and L.M. conceived the analysis, which carried out by H.D.
 H.D., L.M. \& R.H. interpreted the results and wrote the manuscript. 
 \item[Competing Interests] The authors declare that they have no
competing financial interests.
 \item[Correspondence] Correspondence and requests for materials
should be addressed to H.D.(email:hd353@cam.ac.uk)
\end{addendum}

% reference.
% \end{thebibliography}
%\bibliographystylelatex{naturemag}
%\bibliographylatex{reference}
\bibliographystyle{naturemag}
\bibliography{references}

\clearpage
\renewcommand{\figurename}{\textbf{Extended data figure}}

\setcounter{figure}{0}

\begin{figure}
  \centering
  \includegraphics[width=0.8\textwidth]{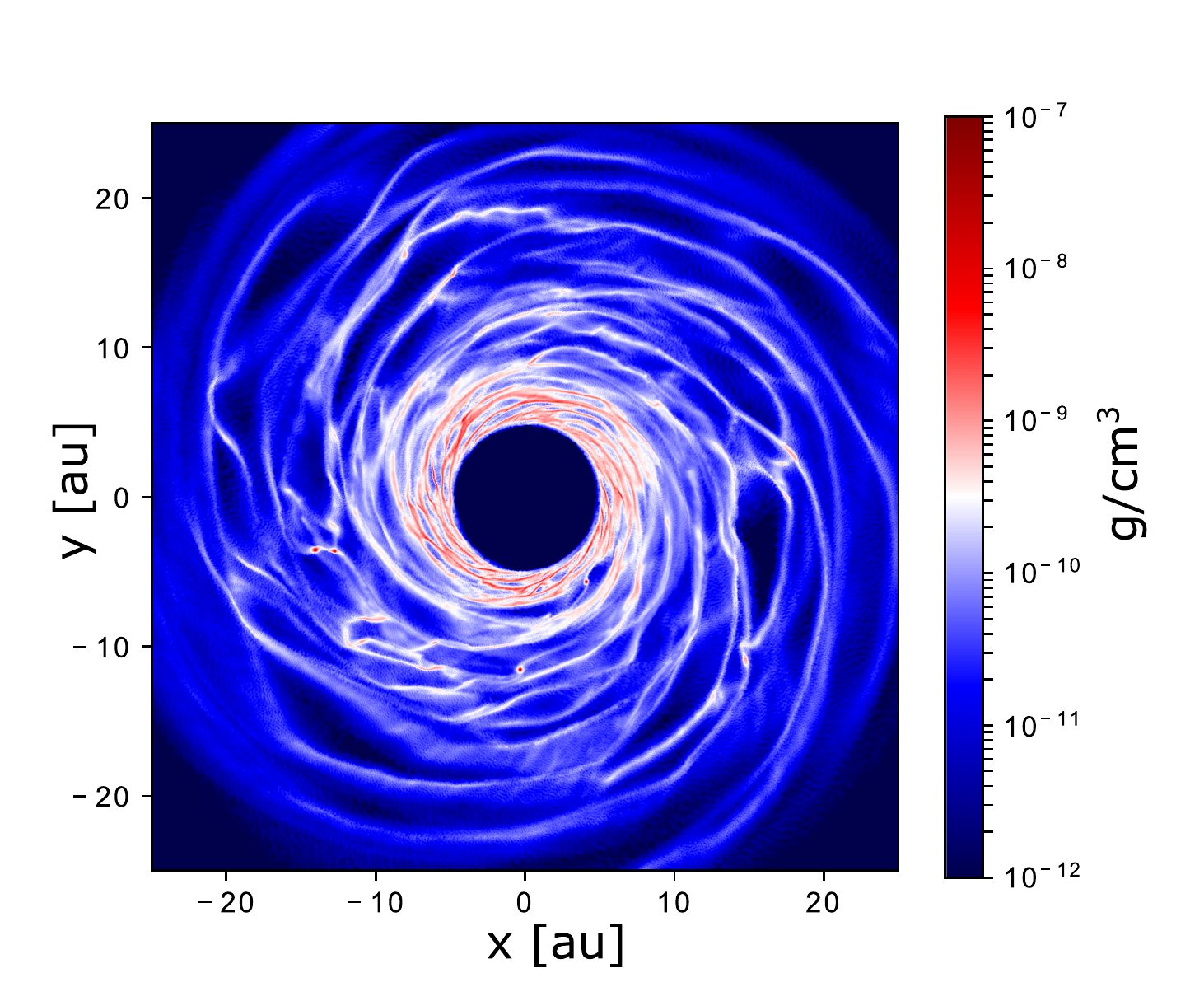}
  \caption{The disk's  midplane density of the HD fragmentation simulation at 100 yrs. Two clumps at 4 and 6 o'clock are formed  with  masses of 0.128$M_J$ and 0.042$M_J$, respectively. The over-dense region near 8 o'clock is not gravitationally bound.}
\end{figure}

\begin{figure}
  \centering
  \includegraphics[width=0.8\textwidth]{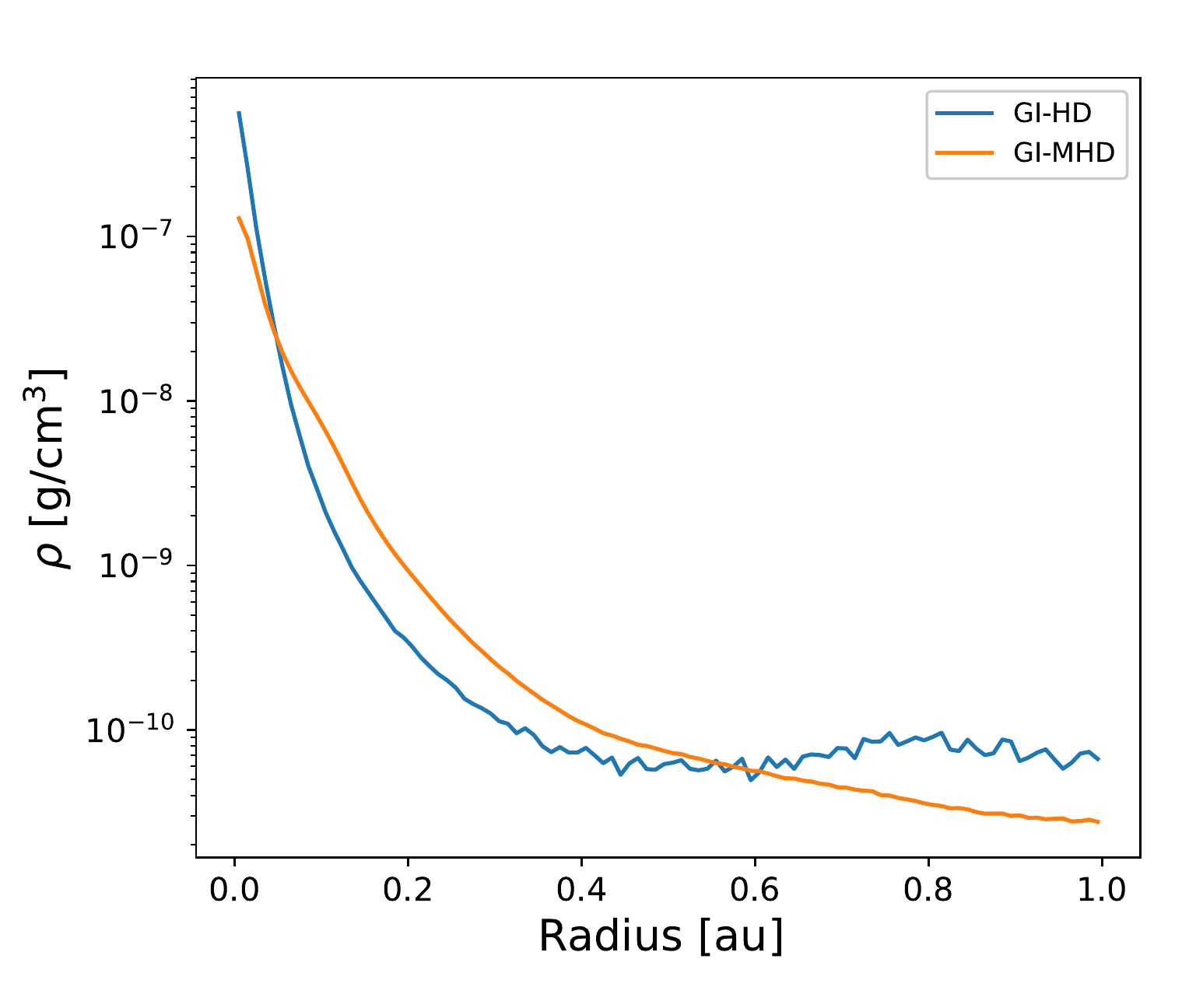}
  \caption{Density profiles around two roughly Neptune-mass protoplanets ($\sim0.04M_J$) in GI-HD clump (HD-cl2 at 100 yrs) and GI-MHD clump (MHD-cl1 at 290 yrs). The cooling time-scale beyond $10^{-9}$g/cm$^{-3}$ region is longer than 50 local orbits. This   justifies turning off cooling in the clump's core region (see Methods subsection 2). However, we note that only regions with a typical density above $\sim 10^{-8}$g/cm$^{-3}$ are gravitationally bound.}
\end{figure}
\begin{figure}
  \centering
  \includegraphics[width=\textwidth]{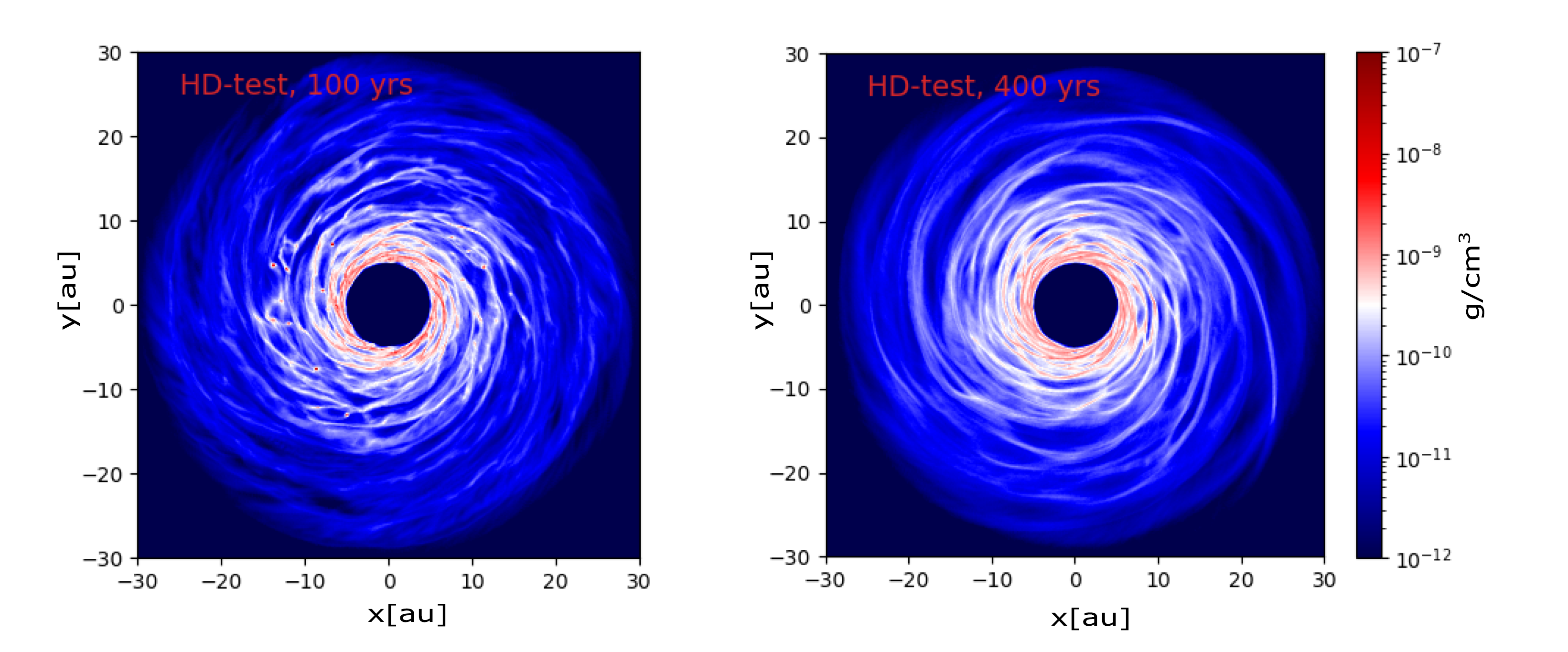}
  \caption{The disk's  density maps in the  resolution test of the HD simulation summarized in Table 1. We employed four times more particles in the test than the HD simulation of Table 1. Also in this case, all the clumps are disrupted after 400 years.}
\end{figure}
\begin{figure}
  \centering
  \includegraphics[width=0.8\textwidth]{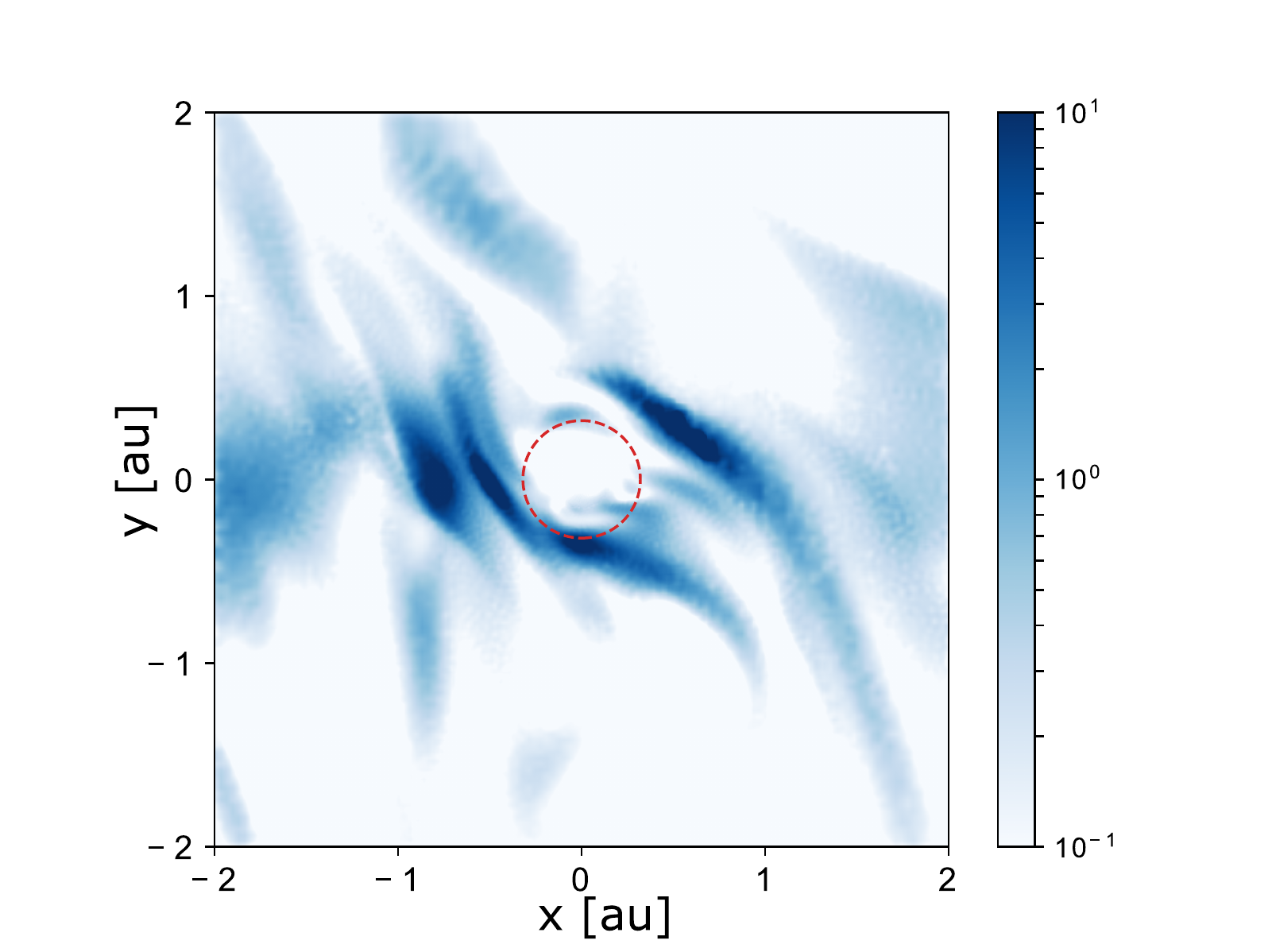}
  \caption{The magnetic energy to kinetic energy ratio in the co-moving frame of MHD-cl1 at 290 yrs (see also the lower panels of Fig.~1). The dashed circle indicates the Hill radius of the protoplanet in MHD-cl1, $~0.3$ au. The magnetic energy dominates the kinetic energy outside the Hill sphere, thus controlling the  dynamics of the flow.}
\end{figure}
\begin{figure}
  \centering
  \includegraphics[width=0.8\textwidth]{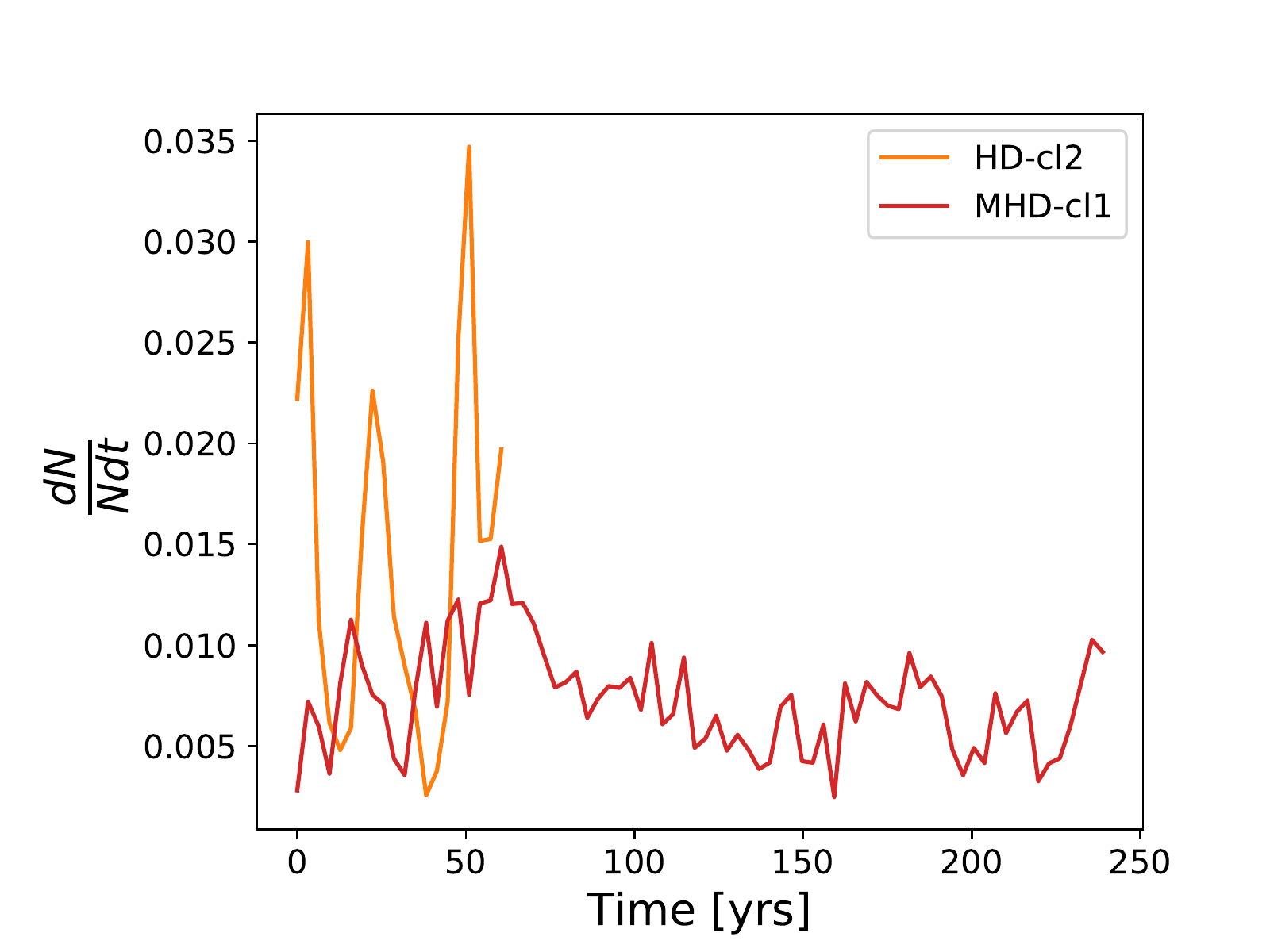}
  \caption{The refilling rate of material within the Hill sphere of a Neptune-mass protoplanet. The refilling rate is defined as the new mass, as a fraction of the total, that enters the Hill sphere of the protoplanet every year. The calculation is performed for HD-cl2 (100-170 yrs) and MHD-cl1 (200-440 yrs)  (see Figure 2) using snapshots taken every 3 yrs. The spikes of the yellow curve are caused by encounters with spiral density waves.}
\end{figure}
\begin{figure}
  \centering
  \includegraphics[width=0.8\textwidth]{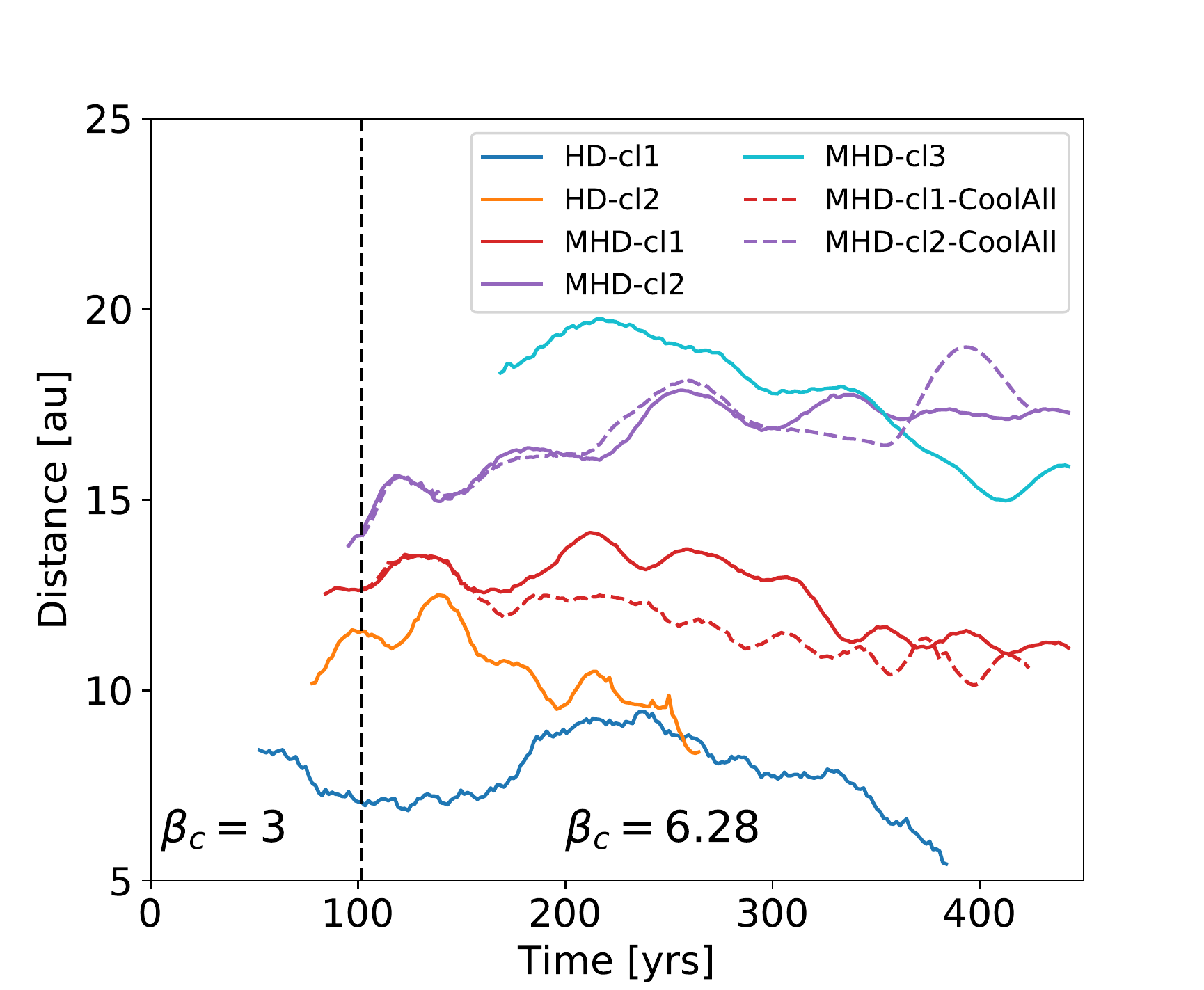}
  \caption{\textbf{Clump migration}. The heliocentric distance of the clumps/protoplanets (see figure 2) suggests both inwards and outwards migration can happen due to gravito-turbulence.}
\end{figure}

%\includepdf[pages=-]{empty.pdf}
%\includepdf[pages=-]{cosmochemistry.pdf}
%\includepdf[pages=-]{Remarks.pdf}
\end{document}